# Robust Parameter Estimation of Regression Model with AR(p) Error Terms

Y. Tuaç, Y. Güney B. Şenoğlu and O. Arslan

Ankara University, Faculty of Science, Department of Statistics, 06100 Ankara/Turkey

ytuac@ankara.edu.tr, ydone@ankara.edu.tr, senoglu@science.ankara.edu.tr, oarslan@ankara.edu.tr

**Abstract**

In this paper, we consider a linear regression model with AR(p) error terms with the assumption that the error terms have a *t* distribution as a heavy tailed alternative to the normal distribution. We obtain the estimators for the model parameters by using the conditional maximum likelihood (CML) method. We conduct an iteratively reweighting algorithm (IRA) to find the estimates for the parameters of interest. We provide a simulation study and three real data examples to illustrate the performance of the proposed robust estimators based on *t* distribution.

**Keywords:** autoregressive stationary process; conditional maximum likelihood; linear regression; non normal distributions; robust estimation.

## 1. Introduction

Consider the following linear regression model

$$y_t = \sum_{i=1}^{M} x_{t,i} \beta_i + e_t, \quad t = 1,2,\dots,N \tag{1}$$

where, $y_t$ is the response variable, $x_{t,i}$ are the explanatory variables, $\beta_i$ are the unknown regression parameters and $e_t$ is the error term. In classical regression analysis, the general assumptions on the error term are zero mean, constant variance and not correlated with each other. It is well known that under these assumptions the ordinary least squares (OLS) estimator is the best. However, one of the problems in application is that the error term may be correlated with each other. In this case, although, the OLS estimators are unbiased and consistence, they may be no longer efficient even in large sample cases, and hence this may cause large estimated standard errors for the estimators of the regression parameters (see Olaomi and Ifederu [13]). There are many ways to deal with autocorrelated structures in the disturbances; the most common way is to assume autoregressive error terms in regression model.

We assume that $e_t$ is a stationary autoregressive error process of order *p* (AR(p)) given as

$$e_t = \phi_1 e_{t-1} + \dots + \phi_p e_{t-p} + a_t, \tag{2}$$

where $\phi_j$, for $j = 1,2,\ldots,p$, are unknown autoregressive parameters.

For simplicity we use $a_t = e_t - \phi_1 e_{t-1} - \cdots - \phi_p e_{t-p} = \Phi(B)e_t$, where $E(a_t) = 0$, $Var(a_t) = \sigma^2$ and $a_t$'s are uncorrelated random variables with constant variance. Here $B$ is called the backshift operator and where $\Phi(\cdot)$ is the function defining the autoregression. Then using the backshift operator the regression model given in (1) can be rewritten as

$$\Phi(B)y_t = \sum_{i=1}^{M} \beta_i \Phi(B)x_{t,i} + a_t, \quad t = p + 1, 2, \ldots, N,$$
(3)

where

$$\Phi(B)y_t = y_t - \phi_1 y_{t-1} - \cdots - \phi_p y_{t-p}, \tag{4}$$
$$\Phi(B)x_{t,i} = x_{t,i} - \phi_1 x_{t-1,i} - \cdots - \phi_p x_{t-p,i} \tag{5}$$

In general, it is assumed that $a_t$ is normally distributed. For instance, Alpuim and El-Shaarawi [2] estimated the parameters of the regression model with AR(p) error term using the OLS estimation method. They also used the maximum likelihood (ML) estimation and CML estimation method under the assumption of normality and studied the asymptotic properties of the resulting estimators. Beach and Mackinnon [5] used ML estimation method to estimate the parameters of AR(1) error term regression models. Tiku [17] estimated the parameters by using the modified maximum likelihood (MML) method for the regression model with AR(1) error terms under the assumption that the error term has the Long Tailed Symmetric (LTS) distribution. There are some other studies used heavy tailed distributions in time series. For instance, Hill [7] used tail-trimming and/or weighting to show how robust to any type of light or heavy tailed distribution in infinite variance autoregressions case. Also, heavy tailed asymmetrically distributed errors in GARCH model were discussed with tail-trimmed QML estimator in Hill [8,9].

Another challenging problem in a regression analysis is the presence of outliers in data. Since the parameter estimators based on normal distribution are very sensitive to the outliers, the corresponding estimators will be no longer efficient. One way to combat with the outliers is to use heavy tailed distributions as alternatives to the normal distribution. Thus, the *t* distribution provides a useful alternative to the normal distribution for statistical modelling of data sets that have heavier tailed empirical distribution. The motivation of this paper is to propose conditional maximum likelihood estimators for unknown parameters of a linear regression model with autoregressive errors under the assumption that the independent identically distributed (iid) error term $a_t$ given in equation (2) has a *t* distribution with known degrees of freedom. The estimators for the parameters of interest obtained under this assumption will be robust in terms of the influence function. It is known that if the degrees of freedom is estimated along with the other parameters the influence function of the resulting estimators will be unbounded and hence they are not going to be robust (Lucas [10]). Therefore, the degrees of freedom is usually taken as fixed and treated as a robustness tuning parameter in robustness studies, for example see Lange et al. [11].

The rest of the paper is organized as follows. In section 2, we first summarize the CML estimation method. Then, we move on the CML estimation for the parameters of regression model with AR(p) error terms under the assumption that $a_t$'s have *t* distribution. We also give the observed Fisher information matrix for the estimators. Note that the observed Fisher information matrix will be used in section 4 to form confidence intervals and to compute the standard errors of the estimators. In section 3, we give an IRA to compute the estimates. A simulation study and three real data examples are given in section 4 to illustrate the performance of the proposed estimators. Finally we conclude the paper with a discussion section.

## 2. Parameters estimation of the AR(p) error term regression model

In this section, since the exact likelihood function could be well approximated by the conditional likelihood function (Ansley [1]) we will first give the CML estimation method. CML estimators are used mainly in cases where ML estimators are difficult to compute. We will briefly give the conditional likelihood estimators under the normality assumption and move on the *t* distribution case. We also provide observed Fisher matrices for both cases.

### 2.1 Conditional likelihood under normality

Let $a_t$'s have the probability density function $f(a_t, \boldsymbol{\theta})$. If we condition on $a_1, a_2, \ldots, a_p$, the conditional log-likelihood function will be

$$lnL = \sum_{t=p+1}^{N} lnf(a_t|a_1, a_2, \ldots, a_{t-p}, \boldsymbol{\theta}). \tag{6}$$

Consider the regression model given in (3). If it is assumed that $a_t$'s are normally distributed the conditional log-likelihood function will be as follows (Alpuim and El-Shaarawi [2]).

$$lnL = c - \frac{N-p}{2} ln\sigma^2 - \frac{1}{2\sigma^2} \sum_{t=p+1}^{N} \left( \Phi(B) y_t - \sum_{i=1}^{M} \beta_i \Phi(B) x_{t,i} \right)^2 \tag{7}$$

Taking the derivatives of the conditional log-likelihood function with respect to unknown parameters and setting to zero yield the following estimating equations.

$$\frac{\partial lnL}{\partial \beta_k} = \frac{1}{\sigma^2} \sum_{t=p+1}^{N} \left( \Phi(B) y_t - \sum_{i=1}^{M} \beta_i \Phi(B) x_{t,i} \right) \Phi(B) x_{t,k} = 0 \tag{8}$$

$$\frac{\partial lnL}{\partial \phi_l} = \frac{1}{\sigma^2} \sum_{t=p+1}^{N} \left( \Phi(B) y_t - \sum_{i=1}^{M} \beta_i \Phi(B) x_{t,i} \right) \left( y_{t-l} - \sum_{i=1}^{M} \beta_i x_{t-l,i} \right) = 0 \tag{9}$$

$$\frac{\partial lnL}{\partial \sigma^2} = -\frac{N-p}{2\sigma^2} + \frac{1}{2\sigma^4} \sum_{t=p+1}^{N} \left( \Phi(B) y_t - \sum_{i=1}^{M} \beta_i \Phi(B) x_{t,i} \right)^2 = 0 \tag{10}$$

Rearranging these equations we get the following estimators.

$$\underline{\hat{\beta}} = \left[\sum_{t=p+1}^{N} \widehat{\Phi}(B)x_t \widehat{\Phi}(B)x_t^T\right]^{-1} \left[\sum_{t=p+1}^{N} \widehat{\Phi}(B)y_t \widehat{\Phi}(B)x_t\right] \tag{11}$$

$$\underline{\hat{\phi}} = R^{-1}(\hat{\beta})R_0(\hat{\beta}) \tag{12}$$

$$\hat{\sigma}^2 = \frac{1}{N-p} \sum_{t=p+1}^{N} \left(\widehat{\Phi}(B)y_t - \underline{\hat{\beta}}\widehat{\Phi}(B)x_t\right)^2 \tag{13}$$

where

$$R_0(\beta) = \begin{bmatrix} \sum_{t=p+1}^{N} e_t e_{t-1} \\ \sum_{t=p+1}^{N} e_t e_{t-2} \\ \vdots \\ \sum_{t=p+1}^{N} e_t e_{t-p} \end{bmatrix}, R(\beta) = \begin{bmatrix} \sum_{t=p+1}^{N} e_{t-1}^2 & \sum_{t=p+1}^{N} e_{t-1} e_{t-2} & \cdots & \sum_{t=p+1}^{N} e_{t-1} e_{t-p} \\ \sum_{t=p+1}^{N} e_{t-2} e_{t-1} & \sum_{t=p+1}^{N} e_{t-2}^2 & \cdots & \sum_{t=p+1}^{N} e_{t-2} e_{t-p} \\ \vdots & \vdots & \ddots & \vdots \\ \sum_{t=p+1}^{N} e_{t-p} e_{t-1} & \sum_{t=p+1}^{N} e_{t-p} e_{t-2} & \cdots & \sum_{t=p+1}^{N} e_{t-p}^2 \end{bmatrix} \tag{14}$$

and $\widehat{\Phi}(B)$ is the backshift operator with the estimates of $\phi_j$.

Using these equations we can rewrite $\underline{\hat{\beta}}$ and $\hat{\sigma}^2$,

$$\underline{\hat{\beta}} = [\widehat{\boldsymbol{\Phi}}(B)\boldsymbol{X}^T \widehat{\boldsymbol{\Phi}}(B)\boldsymbol{X}]^{-1} [\widehat{\boldsymbol{\Phi}}(B)\boldsymbol{X}^T \widehat{\Phi}(B)\underline{Y}] \tag{15}$$

$$\hat{\sigma}^2 = \frac{1}{N-p} [\widehat{\Phi}(B)\underline{Y} - \widehat{\boldsymbol{\Phi}}(B)\boldsymbol{X}\underline{\hat{\beta}}]^T [\widehat{\Phi}(B)\underline{Y} - \widehat{\boldsymbol{\Phi}}(B)\boldsymbol{X}\underline{\hat{\beta}}] \tag{16}$$

where

$\widehat{\boldsymbol{\Phi}}(B)\boldsymbol{X} = [\widehat{\Phi}(B)x_{t,i}],$

$\widehat{\Phi}(B)\underline{Y} = [\widehat{\Phi}(B)y_t].$

These estimators depend on the estimators of the other parameters. Therefore, the values of the estimators should be computed using numerical methods. We use IRA to compute these estimators to guarantee the convergence (see Lange et al. [11], Arslan and Genç [4]). These estimators correspond also to the OLS estimators obtained through the minimization of the sum of squares of the $a_t$. These estimators are sensitive to the outliers in the data. Therefore, an alternative error distribution should be considered to deal with this problem. In the following section we will assume that $a_t$'s have a $t$ distribution with known degrees of freedom and carry out the estimation under this assumption.

Further, we also give the observed Fisher information matrix for the unknown parameters of the regression model defined in equation (3) with normally distributed error terms. Note that the observed Fisher information matrices will be used to compute the standard errors and the confidence intervals in simulation study and the real data examples.

After some straightforward algebra the observed Fisher information matrix for the normal distribution case can be obtained as follows.

$$F(\hat{\beta},\hat{\phi},\hat{\sigma}) = \begin{bmatrix} \frac{1}{\hat{\sigma}^2}(\hat{\Phi}(B)X^T\hat{\Phi}(B)X) & 0 & 0 \\ 0 & \frac{1}{(\hat{\sigma}^2)R(\hat{\beta})} & 0 \\ 0 & 0 & \frac{N-p}{\hat{\sigma}^4} \end{bmatrix} \quad (17)$$

## 2.2 Parameters estimation under $t$ distribution

Consider the regression model given in equation (3) and assume that $a_t$'s have $t$ distribution with the density function

$$f(a_t) = \frac{c_v}{\sigma}\left(v + \frac{a_t^2}{\sigma^2}\right)^{-\frac{v+1}{2}}, \quad (18)$$

where $c_v = \frac{\Gamma\left(\frac{v+1}{2}\right)v^{v/2}}{\sqrt{\pi}\Gamma\left(\frac{v}{2}\right)}$, $v > 0$ degrees of freedom and $\sigma > 0$ scale parameter. Under this assumption the conditional log-likelihood function will be obtained as

$$lnL = lnc_v - (N-p)ln\sigma - \frac{v+1}{2}\sum_{t=p+1}^{N}\ln\left[v + \frac{\left(\Phi(B)y_t - \sum_{i=1}^{M}\beta_i\,\Phi(B)x_{t,i}\right)^2}{\sigma^2}\right]. \quad (19)$$

Taking the derivatives of log-likelihood function with respect to the unknown parameters and setting them to zero yield the following estimating equations.

$$\frac{\partial lnL}{\partial \beta_k} = \frac{(v+1)}{\sigma^2}\sum_{t=p+1}^{N}\frac{\left(\Phi(B)y_t - \sum_{i=1}^{M}\beta_i\,\Phi(B)x_{t,i}\right)\Phi(B)x_{t,k}}{\left[v + \frac{\left(\Phi(B)y_t - \sum_{i=1}^{M}\beta_i\,\Phi(B)x_{t,i}\right)^2}{\sigma^2}\right]} = 0, \quad (20)$$

$$\frac{\partial lnL}{\partial \phi_l} = \frac{(v+1)}{\sigma^2} \sum_{t=p+1}^{N} \frac{\left(\Phi(B)y_t - \sum_{i=1}^{M}\beta_i \Phi(B)x_{t,i}\right)\left(y_{t-l} - \sum_{i=1}^{M}\beta_i x_{t-l,i}\right)}{\left[v + \frac{\left(\Phi(B)y_t - \sum_{i=1}^{M}\beta_i \Phi(B)x_{t,i}\right)^2}{\sigma^2}\right]} = 0, \quad (21)$$

$$\frac{\partial lnL}{\partial \sigma} = -\frac{(N-p)}{\sigma} + \left(\frac{v+1}{\sigma^3}\right) \sum_{t=p+1}^{N} \frac{\left(\Phi(B)y_t - \sum_{i=1}^{M}\beta_i \Phi(B)x_{t,i}\right)^2}{\left[v + \frac{\left(\Phi(B)y_t - \sum_{i=1}^{M}\beta_i \Phi(B)x_{t,i}\right)^2}{\sigma^2}\right]} = 0. \quad (22)$$

Rearranging these equations we get

$$\frac{\partial lnL}{\partial \beta_k} = \frac{1}{\sigma^2} \sum_{t=p+1}^{N} w_t(v) \left(\Phi(B)y_t - \sum_{i=1}^{M}\beta_i \Phi(B)x_{t,i}\right) \Phi(B)x_{t,k} = 0, \quad (23)$$

$$\frac{\partial lnL}{\partial \phi_l} = \frac{1}{\sigma^2} \sum_{t=p+1}^{N} w_t(v) \left(\Phi(B)y_t - \sum_{i=1}^{M}\beta_i \Phi(B)x_{t,i}\right) \left(y_{t-l} - \sum_{i=1}^{M}\beta_i x_{t-l,i}\right) = 0, \quad (24)$$

$$\frac{\partial lnL}{\partial \sigma} = -\frac{(N-p)}{\sigma} + \frac{1}{\sigma^3} \sum_{t=p+1}^{N} w_t(v) \left(\Phi(B)y_t - \sum_{i=1}^{M}\beta_i \Phi(B)x_{t,i}\right)^2 = 0 \quad (25)$$

where

$$w_t(v) = \frac{v+1}{\left[v + \frac{\left(\Phi(B)y_t - \sum_{i=1}^{M}\beta_i \Phi(B)x_{t,i}\right)^2}{\sigma^2}\right]}. \quad (26)$$

These equations yield the following estimators provided that $\left[\sum_{t=p+1}^{N} w_t(v)\widehat{\Phi}(B)x_t\widehat{\Phi}(B)x_t^T\right]^{-1}$ and $R_w^{-1}(\hat{\beta})$ exist.

$$\underline{\hat{\beta}} = \left[\sum_{t=p+1}^{N} w_t(v)\widehat{\Phi}(B)x_t\widehat{\Phi}(B)x_t^T\right]^{-1} \left[\sum_{t=p+1}^{N} w_t(v)\widehat{\Phi}(B)y_t\widehat{\Phi}(B)x_t\right], \quad (27)$$

$$\underline{\hat{\phi}} = R_w^{-1}(\hat{\beta})R_{w0}(\hat{\beta}), \quad (28)$$

$$\hat{\sigma}^2 = \frac{1}{N-p} \sum_{t=p+1}^{N} w_t(v) \left( \hat{\Phi}(B) y_t - \hat{\Phi}(B) x_t^T \underline{\hat{\beta}} \right)^2 \tag{29}$$

where

$$R_{w0}(\beta) = \begin{bmatrix} \sum_{t=p+1}^{N} w_t e_t e_{t-1} \\ \sum_{t=p+1}^{N} w_t e_t e_{t-2} \\ \vdots \\ \sum_{t=p+1}^{N} w_t e_t e_{t-p} \end{bmatrix}, \quad R_w(\beta) = \begin{bmatrix} \sum_{t=p+1}^{N} w_t e_{t-1}^2 & \sum_{t=p+1}^{N} w_t e_{t-1} e_{t-2} & \cdots & \sum_{t=p+1}^{N} w_t e_{t-1} e_{t-p} \\ \sum_{t=p+1}^{N} w_t e_{t-2} e_{t-1} & \sum_{t=p+1}^{N} w_t e_{t-2}^2 & \cdots & \sum_{t=p+1}^{N} w_t e_{t-2} e_{t-p} \\ \vdots & \vdots & \ddots & \vdots \\ \sum_{t=p+1}^{N} w_t e_{t-p} e_{t-1} & \sum_{t=p+1}^{N} w_t e_{t-p} e_{t-2} & \cdots & \sum_{t=p+1}^{N} w_t e_{t-p}^2 \end{bmatrix}. \tag{30}$$

Further, these equations can be rewritten as

$$\underline{\hat{\beta}} = \left[ \hat{\Phi}(B) X^T W \hat{\Phi}(B) X \right]^{-1} \left[ \hat{\Phi}(B) X^T W \hat{\Phi}(B) \underline{Y} \right], \tag{31}$$

$$\hat{\sigma}^2 = \frac{1}{N-p} \left[ \hat{\Phi}(B) \underline{Y} - \hat{\Phi}(B) X \underline{\hat{\beta}} \right]^T W \left[ \hat{\Phi}(B) \underline{Y} - \hat{\Phi}(B) X \underline{\hat{\beta}} \right], \tag{32}$$

by using vector notation. Here

$$\hat{\Phi}(B) X = \left[ \hat{\Phi}(B) x_{t,i} \right]_{\substack{t=p+1,\dots,N \\ i=1,\dots,M}},$$
$$\hat{\Phi}(B) \underline{Y} = \left[ \hat{\Phi}(B) y_t \right]_{t=p+1,\dots,N},$$
$$W = diag\{w_t\}_{t=p+1,\dots,N}.$$

Since the weight function $w_t$ is a decreasing function of $\frac{\left( \Phi(B) y_t - \sum_{i=1}^{M} \beta_i \Phi(B) x_{t,i} \right)^2}{\sigma^2}$ the observations with larger residuals receive small weights. Therefore, the effect of the corresponding point on the estimator will be downweighted. This behavior of the $t$ distribution guarantees the robustness of the resulting estimators (Lucas [10], Arslan and Genç [3, 4]). Note that as $v$ tends to infinity $w_t(v) \to 1$ and this case gives the estimators given in equations (11)-(13).

Similarly the observed Fisher information matrix for the unknown parameters of the regression model defined in equation (3) with the $t$ distributed error terms can be obtained as follows.

$$F(\hat{\beta},\hat{\phi},\hat{\sigma}) = \begin{bmatrix} \frac{1}{\hat{\sigma}^2}(\hat{\Phi}(B)X^TW\hat{\Phi}(B)X) & 0 & 0 \\ 0 & \frac{1}{(\hat{\sigma}^2)R_w(\hat{\beta})} & 0 \\ 0 & 0 & \frac{N-p}{\hat{\sigma}^4} \end{bmatrix} \qquad (33)$$

We should also note that since the estimators given in (27)-(29) are dependent on the weights and since the weights are also functions of the estimators these equations cannot be solved explicitly. Therefore, the numerical methods are also needed to solve these equations to get the estimates. Because of the form of the equations the IRA can be easily implemented to get the estimates as it is done for all the procedures based on $t$ distribution. Note that in the $t$ distribution case the IRA is an expectation-maximization (EM) algorithm so that its convergence is guaranteed (see Lange et al. [11], McLachlan and Krishnan, [12] Arslan and Genç [4]). The following section is devoted to the IRA.

### 3. Iteratively reweighted algorithm

Using the updating equations (28, 31, 32) and the weight function given in equation (26) the following iteratively reweighted algorithm can be formed to calculate the estimates for, $\beta$, $\phi$ and $\sigma^2$. Note that the degrees of freedom of the $t$ distribution will be taken as known and fixed.

(i) Set the initial values $\beta^{(0)}, \phi^{(0)}$ and $\sigma^{2(0)}$ and fix a stopping rule $\delta$.

(ii) Calculate the following weight function for $m = 0,1,2 \dots$

$$w_t^{(m)} = \frac{v+1}{\left[ v + \frac{\left( \Phi^{(m)}(B)y_t - \sum_{i=1}^{M} \beta_i^{(m)} \Phi^{(m)}(B)x_{t,i} \right)^2}{\sigma^{2(m)}} \right]}. \qquad (34)$$

(iii) Using these values calculate

$$\underline{\phi}^{(m+1)} = R_w^{-1(m)}(\hat{\beta}^{(m)})R_{w0}^{(m)}(\hat{\beta}^{(m)}). \qquad (35)$$

(iv) Using $w_t^{(m)}$ and $\underline{\phi}^{(m+1)}$ calculate

$$\underline{\beta}^{(m+1)} = \left[ \Phi^{(m+1)}(B)X^TW^{(m)}\Phi^{(m+1)}(B)X \right]^{-1} \left[ \Phi^{(m+1)}(B)X^TW^{(m)}\Phi^{(m+1)}(B)\underline{Y} \right]. \qquad (36)$$

(v) Using $w_t^{(m)}$, $\underline{\phi}^{(m+1)}$ and $\underline{\beta}^{(m+1)}$ calculate

$$(\sigma^2)^{(m+1)} = \frac{1}{N-p} \left[ \Phi^{(m+1)}(B)\underline{Y} - \mathbf{\Phi}^{(m+1)}(\mathbf{B})\mathbf{X}\underline{\beta}^{(m+1)} \right]^T W^{(m)} \left[ \Phi^{(m+1)}(B)\underline{Y} - \mathbf{\Phi}^{(m+1)}(\mathbf{B})\mathbf{X}\underline{\beta}^{(m+1)} \right] \quad (37)$$

(vi) Repeat the steps (ii)-(v) until the convergence condition $max(\|\beta^{(m+1)} - \beta^{(m)}\|, \|\phi^{(m+1)} - \phi^{(m)}\|, \|(\sigma^2)^{(m+1)} - (\sigma^2)^{(m)}\|) < \delta$ is satisfied.

In section 4, we will use this algorithm to compute the CML estimates in simulation and the real data examples.

## 4. Numerical study

In this section, we give a small simulation study and three real data examples to illustrate the performance of the regression estimators obtained from the *t* distribution (with AR (2) error terms for finite sample case) with and without outliers in the data. The CML estimates are computed using the IRA given in section 3. Note that throughout the simulation study and the real data examples the degrees of freedom ($v$) of the *t* distribution is taken as 3 since the small values such as 3 are suggested for the sake of robustness in literature (e.g see Lange et.al, 1989).

### 4.1 A simulation study

*Simulation design.* Firstly we generate three independent variables $x_t$ from standard normal distribution ($x_{t,i} \sim N(0,1)$). The values of the parameters are $\underline{\beta} = (\beta_1, \beta_2, \beta_3)' = (0.1, 0.5, 0.9)'$ and $\underline{\phi} = (\phi_1, \phi_2)' = (-0.7, 0.12)'$. Note that the AR (2) model values of $\underline{\phi}$ are taken to guarantee the stationarity assumption for the model of the error terms. Then the values of the response variable are generated using $\Phi(B)y_t = \sum_{i=1}^{M} \beta_i \Phi(B) x_{t,i} + a_t$.

*Simulation cases.* In first case the $a_t$'s in (3) are generated from standard normal distribution ($a_t \sim N(0,1)$) and parameters are estimated by using normal distribution's conditional likelihood and *t* distribution's conditional likelihood. In the second case the $a_t$'s are generated from *t* distribution with $v = 3$ degrees of freedom ($a_t \sim t_v(0,1)$) and parameters are estimated by using normal and *t* assumptions again. The last case is for the symmetric Pareto distributed error terms, where we use the Pareto distribution symmetric by zero as given in [6] with the tail index (shape) parameter $\kappa = 1.25$. For the Pareto error case we compute the values of the estimators obtained from the normal and the *t* distributions.

*Outlier case.* To add some outliers to the data 10 percent of $\underline{Y}$ is replaced by the points generated from $N(0, 100)$.

*Performance measures.* Mean squared error (MSE) and bias values are calculated to compare the estimators. These values are calculated by using $R = 100$ replications for the sample sizes $n = 25, 50$ and 100. Also, using the observed Fisher information matrix given in section 2, the standard errors (SE) and the confidence intervals (CIL - CIU) are calculated.

The MSE values and the biases are calculated using

$$MSE(\hat{\beta}) = \frac{1}{R}\sum_{i=1}^{R}(\hat{\beta}_i - \beta)^2, \; bias(\hat{\beta}) = \bar{\beta} - \beta,$$

$$MSE(\hat{\phi}) = \frac{1}{R}\sum_{i=1}^{R}(\hat{\phi}_i - \phi)^2, \; bias(\hat{\phi}) = \bar{\phi} - \phi,$$

$$MSE(\hat{\sigma}^2) = \frac{1}{R}\sum_{i=1}^{R}(\hat{\sigma}_i^2 - \sigma^2)^2, \; bias(\hat{\sigma}^2) = \overline{\sigma^2} - \sigma^2,$$

where $\bar{\beta} = \frac{1}{R}\sum_{i=1}^{R}\hat{\beta}_i$, $\bar{\phi} = \frac{1}{R}\sum_{i=1}^{R}\hat{\phi}_i$, $\overline{\sigma^2} = \frac{1}{R}\sum_{i=1}^{R}\hat{\sigma}_i^2$.

*Simulation results.* Simulation results are given in Tables 1 and 2.

*Simulation results without outliers*

Table 1. Bias, MSE, SE, CIL and CIU values of the estimates without outliers
True Values $(\beta_1, \beta_2, \beta_3)' = (0.1, 0.5, 0.9)'$, $(\phi_1, \phi_2)' = (-0.7, 0.12)$ and $\sigma = 1$

| n | | | Normal Error | | t Error | | Paretian Error ($\kappa = 1.25$) | |
|---|---|---|---|---|---|---|---|---|
| | | | Norm | St-t | Norm | St-t | Norm | St-t |
| 25 | | $\hat{\beta}_1$ | 0.0854 | 0.0749 | 0.1135 | 0.1360 | 25.3813 | -0.2909 |
| | | Bias | 0.0230 | 0.0302 | -0.0136 | -0.0075 | 25.2813 | -0.3909 |
| | $\beta_1$ | MSE | 0.0453 | 0.0578 | 0.0789 | 0.0708 | 26940.5 | 20.0117 |
| | | SE | 0.1598 | 0.3965 | 0.2678 | 0.5131 | 5.0918 | 0.7380 |
| | | CIL | 0.0501 | -0.0366 | 0.0082 | 0.0272 | 23.4241 | -0.5746 |
| | | CIU | 0.1754 | 0.2484 | 0.2189 | 0.2449 | 27.3385 | -0.0073 |
| | | $\hat{\beta}_2$ | 0.4817 | 0.4717 | 0.4945 | 0.4600 | 24.2895 | -0.0754 |
| | | Bias | 0.0182 | -0.0293 | -0.0263 | 0.0086 | 23.6061 | -0.5754 |
| | $\beta_2$ | MSE | 0.0372 | 0.0656 | 0.0945 | 0.1150 | 61730.1 | 56.5089 |
| | | SE | 0.2119 | 0.2997 | 0.2235 | 0.4930 | 5.4498 | 1.8644 |
| | | CIL | 0.4044 | 0.3513 | 0.4045 | 0.3677 | 22.1946 | -0.7921 |
| | | CIU | 0.6008 | 0.6417 | 0.5846 | 0.5522 | 26.3843 | 0.6412 |
| | | $\hat{\beta}_3$ | 0.8959 | 0.9039 | 0.8800 | 0.9008 | -20.515 | 0.6152 |
| | | Bias | 0.0128 | 0.0130 | 0.0112 | -0.0125 | -21.415 | -0.2848 |
| | $\beta_3$ | MSE | 0.0468 | 0.0537 | 0.0901 | 0.1110 | 17290.3 | 10.3356 |
| | | SE | 0.1813 | 0.3596 | 0.3182 | 0.4130 | 2.9110 | 0.4780 |
| | | CIL | 0.8266 | 0.7264 | 0.7574 | 0.7370 | -21.634 | 0.4315 |
| | | CIU | 0.9687 | 1.0499 | 1.0025 | 1.0645 | -19.396 | 0.7989 |
| | | $\hat{\phi}_1$ | -0.7034 | -0.5273 | -0.7484 | -0.5611 | -0.5226 | -0.5219 |
| | | Bias | -0.0326 | 0.0923 | -0.0240 | 0.0818 | 0.1774 | 0.2995 |
| | $\phi_1$ | MSE | 0.0715 | 0.0810 | 0.0699 | 0.0618 | 0.0973 | 0.1307 |
| | | SE | 0.0261 | 0.1414 | 0.0826 | 0.2476 | 0.1617 | 0.1971 |
| | | CIL | -0.7136 | -0.6208 | -0.7833 | -0.6007 | -0.5845 | -0.4763 |
| | | CIU | -0.6931 | -0.4338 | -0.7136 | -0.5214 | -0.4606 | -0.3247 |
| | | $\hat{\phi}_2$ | 0.0276 | 0.0626 | 0.0283 | 0.0413 | 0.2662 | 0.2888 |
| | | Bias | -0.1222 | -0.1412 | -0.1013 | -0.1295 | 0.1462 | 0.1251 |
| | $\phi_2$ | MSE | 0.0832 | 0.0854 | 0.0825 | 0.0669 | 0.0857 | 0.0404 |
| | | SE | 0.1403 | 0.1397 | 0.1227 | 0.1835 | 0.9441 | 0.1710 |
| | | CIL | -0.0644 | -0.0111 | -0.0213 | -0.0027 | -0.0987 | 0.1794 |
| | | CIU | 0.1196 | 0.1364 | 0.0778 | 0.0852 | 0.6310 | 0.3108 |

|  |  |  | Normal Error | | t Error | | Paretian Error ($\kappa = 1.25$) | |
|---|---|---|---|---|---|---|---|---|
|  |  |  | Norm | St-t | Norm | St-t | Norm | St-t |
|  | $\sigma$ | $\hat{\sigma}$ | 0.8711 | 0.6923 | 1.3911 | 0.8149 | 169.849 | 0.9941 |
|  |  | Bias | -0.1101 | 0.3134 | 0.4137 | 0.2423 | 168.849 | -0.0059 |
|  |  | MSE | 0.0349 | 0.1105 | 0.3948 | 0.1188 | 652360 | 0.2501 |
|  |  | SE | 0.1241 | 0.0963 | 0.2948 | 0.1381 | 34.675 | 0.2017 |
|  |  | CIL | 0.7894 | 0.6490 | 1.2774 | 0.7606 | 156.522 | 0.9166 |
|  |  | CIU | 0.9300 | 0.7264 | 1.5048 | 0.8692 | 183.176 | 1.0717 |
|  |  |  | Normal Error | | t Error | | Paretian Error ($\kappa = 1.25$) | |
|  |  |  | Norm | St-t | Norm | St-t | Norm | St-t |
| 50 | $\beta_1$ | $\hat{\beta}_1$ | 0.0863 | 0.0792 | 0.0677 | 0.0883 | -2.9119 | 0.3695 |
|  |  | Bias | -0.0137 | -0.0208 | -0.0323 | -0.0117 | -3.0119 | -0.2695 |
|  |  | MSE | 0.0198 | 0.0258 | 0.0557 | 0.0399 | 499.264 | 4.5549 |
|  |  | SE | 0.3551 | 0.2667 | 0.2473 | 0.3190 | 1.8167 | 0.3675 |
|  |  | CIL | -0.0121 | 0.0229 | 0.0069 | 0.0252 | -3.4155 | 0.2693 |
|  |  | CIU | 0.1847 | 0.1555 | 0.1346 | 0.1513 | -2.4083 | 0.4697 |
|  | $\beta_2$ | $\hat{\beta}_2$ | 0.5004 | 0.4949 | 0.5349 | 0.5310 | -1.0789 | 0.3599 |
|  |  | Bias | 0.0004 | -0.0051 | -0.0349 | 0.0310 | -1.5789 | -0.1401 |
|  |  | MSE | 0.0118 | 0.0163 | 0.0401 | 0.0307 | 353.475 | 2.4658 |
|  |  | SE | 0.1746 | 0.2404 | 0.2325 | 0.2985 | 0.9547 | 0.9719 |
|  |  | CIL | 0.4520 | 0.4787 | 0.4687 | 0.4448 | -1.3435 | 0.0898 |
|  |  | CIU | 0.5488 | 0.5112 | 0.6010 | 0.6172 | -0.8142 | 0.6300 |
|  | $\beta_3$ | $\hat{\beta}_3$ | 0.8986 | 0.9083 | 0.9122 | 0.9088 | 1.7127 | 1.1110 |
|  |  | Bias | -0.0014 | 0.0083 | -0.0122 | 0.0088 | -2.6127 | 0.2110 |
|  |  | MSE | 0.0186 | 0.0196 | 0.0496 | 0.0310 | 792.217 | 1.4357 |
|  |  | SE | 0.2209 | 0.1215 | 0.1092 | 0.2660 | 2.0044 | 1.3471 |
|  |  | CIL | 0.8363 | 0.8831 | 0.8832 | 0.8802 | -2.2683 | 0.7256 |
|  |  | CIU | 0.9610 | 0.9531 | 0.9411 | 0.9374 | -1.1572 | 1.4965 |
|  | $\phi_1$ | $\hat{\phi}_1$ | -0.7449 | -0.6906 | -0.6860 | -0.5060 | -0.5743 | -0.4243 |
|  |  | Bias | -0.0449 | 0.1968 | 0.0140 | 0.1932 | 0.1257 | 0.2757 |
|  |  | MSE | 0.0264 | 0.0431 | 0.0273 | 0.0567 | 0.0426 | 0.1184 |
|  |  | SE | 0.2443 | 0.0962 | 0.0974 | 0.1196 | 0.0725 | 0.3794 |
|  |  | CIL | -0.8122 | -0.8290 | -0.7127 | -0.5345 | -0.6029 | -0.5284 |
|  |  | CIU | -0.6775 | -0.5575 | -0.6593 | -0.4747 | -0.5627 | -0.3202 |
|  | $\phi_2$ | $\hat{\phi}_2$ | 0.0543 | 0.0803 | 0.0788 | 0.0945 | 0.1740 | 0.1980 |
|  |  | Bias | -0.0657 | -0.0441 | -0.0412 | -0.0896 | 0.1078 | 0.0780 |
|  |  | MSE | 0.0309 | 0.0206 | 0.0279 | 0.0199 | 0.0347 | 0.0210 |
|  |  | SE | 0.1643 | 0.1237 | 0.1146 | 0.1161 | 0.0825 | 0.3210 |
|  |  | CIL | 0.0076 | 0.0390 | 0.0464 | 0.0559 | 0.1512 | 0.1090 |
|  |  | CIU | 0.1010 | 0.1127 | 0.1112 | 0.1250 | 0.1969 | 0.2871 |
|  | $\sigma$ | $\hat{\sigma}$ | 0.9334 | 0.7397 | 1.5421 | 0.9104 | 60.540 | 1.1750 |
|  |  | Bias | -0.0666 | -0.2603 | 0.5421 | -0.0896 | 59.540 | 0.1750 |
|  |  | MSE | 0.0170 | 0.0742 | 0.4703 | 0.0199 | 25203 | 0.3459 |
|  |  | SE | 0.1347 | 0.0836 | 0.2265 | 0.1157 | 8.7382 | 0.1975 |
|  |  | CIL | 0.8960 | 0.7178 | 1.4804 | 0.8772 | 58.1179 | 1.1198 |
|  |  | CIU | 0.9707 | 0.7616 | 1.6038 | 0.9435 | 65.9622 | 1.2303 |
|  |  |  | Normal Error | | t Error | | Paretian Error ($\kappa = 1.25$) | |
|  |  |  | Norm | St-t | Norm | St-t | Norm | St-t |
| 100 | $\beta_1$ | $\hat{\beta}_1$ | 0.1001 | 0.0964 | 0.1288 | 0.1085 | 4.9025 | 0.1551 |
|  |  | Bias | 0.0001 | -0.0036 | 0.0288 | 0.0085 | 4.8025 | 0.0551 |
|  |  | MSE | 0.0072 | 0.0106 | 0.0186 | 0.0103 | 2028.2 | 0.7551 |

|  |  |  |  |  |  |  |  |
|---|---|---|---|---|---|---|---|
|  | SE | 0.1271 | 0.1666 | 0.2197 | 0.1995 | 3.0280 | 0.3802 |
|  | CIL | 0.0805 | 0.0769 | 0.0929 | 0.0653 | 4.2516 | 0.0596 |
|  | CIU | 0.1197 | 0.1159 | 0.1647 | 0.1516 | 5.5535 | 0.3095 |
|  | $\hat{\beta}_2$ | 0.5123 | 0.5129 | 0.5054 | 0.4990 | 19.0216 | 0.3903 |
|  | Bias | 0.0123 | 0.0129 | 0.0054 | -0.0010 | 18.5216 | -0.1097 |
| $\beta_2$ | MSE | 0.0078 | 0.0101 | 0.0186 | 0.0135 | 9080.24 | 1.5062 |
|  | SE | 0.1415 | 0.1827 | 0.2314 | 0.2146 | 3.0746 | 0.4928 |
|  | CIL | 0.4969 | 0.4885 | 0.4890 | 0.4632 | 18.3705 | 0.1974 |
|  | CIU | 0.5277 | 0.5372 | 0.5217 | 0.5348 | 19.6726 | 0.5832 |
|  | $\hat{\beta}_3$ | 0.8996 | 0.9047 | 0.9211 | 0.9197 | 20.3116 | 0.7991 |
|  | Bias | -0.0004 | 0.0047 | 0.0211 | 0.0197 | 19.4116 | -0.1009 |
| $\beta_3$ | MSE | 0.0062 | 0.0083 | 0.0174 | 0.0146 | 3555.75 | 0.6635 |
|  | SE | 0.1340 | 0.1736 | 0.2144 | 0.1940 | 3.1505 | 0.2957 |
|  | CIL | 0.8745 | 0.8754 | 0.8888 | 0.8853 | 19.8952 | 0.7643 |
|  | CIU | 0.9248 | 0.9341 | 0.9534 | 0.9541 | 20.7280 | 0.8339 |
|  | $\hat{\phi}_1$ | -0.7217 | -0.5309 | -0.6996 | -0.5160 | -0.6227 | -0.3619 |
|  | Bias | -0.0217 | 0.1691 | 0.0004 | 0.1840 | 0.0773 | 0.3381 |
| $\phi_1$ | MSE | 0.0148 | 0.0389 | 0.0091 | 0.0425 | 0.0188 | 0.1393 |
|  | SE | 0.0881 | 0.0795 | 0.1357 | 0.0807 | 0.6589 | 0.7619 |
|  | CIL | -07408 | -0.5467 | -0.7180 | -0.5330 | -0.6344 | -0.3768 |
|  | CIU | -07026 | -0.5150 | -0.6812 | -0.4990 | -0.6110 | -0.3469 |
|  | $\hat{\phi}_2$ | 0.0912 | 0.0980 | 0.1072 | 0.0973 | 0.1961 | 0.2485 |
|  | Bias | -0.0288 | -0.0220 | -0.0128 | -0.0227 | 0.0761 | 0.1285 |
| $\phi_2$ | MSE | 0.0161 | 0.0101 | 0.0110 | 0.0070 | 0.0161 | 0.0276 |
|  | SE | 0.0882 | 0.0772 | 0.1359 | 0.0724 | 1.0997 | 0.7410 |
|  | CIL | 0.0693 | 0.0797 | 0.0868 | 0.0807 | 0.1826 | 0.2337 |
|  | CIU | 0.1131 | 0.1162 | 0.1276 | 0.1139 | 0.2096 | 0.2632 |
|  | $\hat{\sigma}$ | 0.9748 | 0.7721 | 1.5719 | 0.9178 | 175.337 | 1.2211 |
|  | Bias | -0.0252 | -0.2279 | 0.5719 | -0.0822 | 174.332 | 0.2211 |
| $\sigma$ | MSE | 0.0064 | 0.0550 | 0.4113 | 0.0126 | 444540 | 0.3115 |
|  | SE | 0.0891 | 0.0612 | 0.1660 | 0.0843 | 16.5886 | 0.1521 |
|  | CIL | 0.9555 | 0.7603 | 1.5407 | 0.9012 | 191.887 | 1.1916 |
|  | CIU | 0.9941 | 0.7839 | 1.6030 | 0.9345 | 159.144 | 1.2507 |

Table 1 displays the simulation results for the case without outlier in the data with different sample sizes. From the table we can say that error terms based on normal distribution and $t$ distribution cases have similar performance. When the error distribution is normal the CML estimators based on the normal distribution perform the best and the performance of the CML estimator based on the $t$ distribution is comparable with the estimators based on the normal distribution. On the other hand, if the error distribution is the $t$ distribution the estimators based on $t$ distribution are the best, and it is followed by the normal distribution. Finally, for the Pareto distributed error case the estimators based on the normal distribution drastically affected and give the worst results with the larger MSE and the bias values. But, for the Pareto distributed error case the estimators obtained from the $t$ distribution behave much better than the normal case.

*Simulation results with 10% outliers*

Table 2. Bias, MSE, SE, CIL and CIU values of the estimates with outliers
True Values $(\beta_1, \beta_2, \beta_3)' = (0.1, 0.5, 0.9)'$, $(\phi_1, \phi_2)' = (-0.7, 0.12)$ and $\sigma = 1$

| n | Parameter | | Normal Error | | t Error | |
|---|---|---|---|---|---|---|
| | | | Normal | t | Normal | t |
| 25 | $\beta_1$ | $\hat{\beta}_1$ | -0.1805 | 0.1277 | -0.7210 | -0.0001 |
| | | Bias | -0.2805 | 0.0277 | -0.8210 | -0.1001 |
| | | MSE | 14.395 | 0.0666 | 24.5021 | 0.1529 |
| | | SE | 0.7206 | 0.6763 | 0.6216 | 0.6371 |
| | | CIL | -0.3517 | -0.2601 | -0.8925 | -0.2184 |
| | | CIU | -0.0094 | 0.5154 | -0.5494 | 0.2182 |
| | $\beta_2$ | $\hat{\beta}_2$ | 1.1081 | 0.5161 | 0.4523 | 0.5571 |
| | | Bias | 0.6081 | 0.0161 | -0.0477 | 0.0571 |
| | | MSE | 19.419 | 0.1170 | 14.6355 | 0.2807 |
| | | SE | 2.2292 | 0.7016 | 0.6484 | 0.5687 |
| | | CIL | 0.9673 | 0.1159 | 0.2394 | 0.2805 |
| | | CIU | 1.2490 | 0.9163 | 0.6652 | 0.8336 |
| | $\beta_3$ | $\hat{\beta}_3$ | 0.2261 | 0.9344 | 0.6559 | 0.9139 |
| | | Bias | -0.6739 | 0.0344 | -0.2441 | 0.0139 |
| | | MSE | 15.333 | 0.0984 | 18.1168 | 0.1785 |
| | | SE | 2.2848 | 0.8055 | 0.9610 | 0.9656 |
| | | CIL | 0.0442 | 0.7948 | 0.5443 | 0.7959 |
| | | CIU | 0.4079 | 1.0740 | 0.7674 | 1.0320 |
| | $\phi_1$ | $\hat{\phi}_1$ | -0.0335 | -0.2303 | 0.0294 | -0.2678 |
| | | Bias | 0.6665 | 0.4697 | 0.7294 | 0.4322 |
| | | MSE | 3.0020 | 0.3129 | 2.1621 | 0.3226 |
| | | SE | 5.2471 | 0.0421 | 0.3258 | 0.0030 |
| | | CIL | -0.2352 | -0.3086 | -0.1113 | -0.3572 |
| | | CIU | 0.1683 | -0.1521 | 0.1701 | -0.1784 |
| | $\phi_2$ | $\hat{\phi}_2$ | -2.0472 | 0.1866 | -1.7184 | 0.2123 |
| | | Bias | -2.1672 | 0.0666 | -1.8384 | 0.1005 |
| | | MSE | 19.1661 | 0.0970 | 13.2772 | 0.0923 |
| | | SE | 2.2688 | 0.0995 | 0.3081 | 0.1152 |
| | | CIL | -2.1668 | 0.0904 | -2.2067 | 0.1214 |
| | | CIU | -1.9275 | 0.2829 | -1.2301 | 0.3031 |
| | $\sigma$ | $\hat{\sigma}$ | 17.8423 | 0.9916 | 19.3621 | 1.1540 |
| | | Bias | 16.8423 | -0.0084 | 18.3621 | 0.1540 |
| | | MSE | 417.615 | 0.0277 | 452.471 | 0.0759 |
| | | SE | 87.2684 | 0.1956 | 4.2045 | 0.2637 |
| | | CIL | 16.3839 | 0.9112 | 17.7795 | 1.0451 |
| | | CIU | 19.3007 | 1.0720 | 20.9447 | 1.2628 |
| | | | Normal Error | | t Error | |
| | | | Normal | t | Normal | t |
| 50 | $\beta_1$ | $\hat{\beta}_1$ | -0.2296 | 0.1139 | -0.2276 | 0.0913 |
| | | Bias | -0.3296 | 0.0139 | -0.3276 | -0.0087 |
| | | MSE | 11.890 | 0.0363 | 9.6600 | 0.0531 |
| | | SE | 30.410 | 0.4740 | 2.9891 | 0.7061 |
| | | CIL | -0.8091 | -0.0874 | -0.6726 | -0.0615 |
| | | CIU | 0.3500 | 0.3152 | 0.2173 | 0.2440 |

|   |   |   | Normal Error | | $t$ Error | |
|---|---|---|---|---|---|---|
|   |   |   | Normal | $t$ | Normal | $t$ |
|   | $\beta_2$ | $\hat{\beta}_2$ | 0.9483 | 0.4933 | 0.3304 | 0.5128 |
|   |   | Bias | 0.4483 | -0.0067 | -0.1696 | 0.0128 |
|   |   | MSE | 14.957 | 0.0228 | 13.637 | 0.0557 |
|   |   | SE | 25.740 | 0.3973 | 2.8076 | 0.4466 |
|   |   | CIL | 0.3299 | 0.3797 | -0.2893 | 0.3497 |
|   |   | CIU | 1.5667 | 0.6069 | 0.9501 | 0.6758 |
|   | $\beta_3$ | $\hat{\beta}_3$ | 0.4665 | 0.8961 | 0.9969 | 0.9010 |
|   |   | Bias | -0.4335 | -0.0039 | 0.0969 | 0.0010 |
|   |   | MSE | 16.155 | 0.0307 | 10.6562 | 0.0540 |
|   |   | SE | 31.805 | 0.2983 | 1.4603 | 0.7257 |
|   |   | CIL | 0.1967 | 0.7335 | 0.5512 | 0.8402 |
|   |   | CIU | 0.7363 | 1.0587 | 1.4426 | 0.9619 |
|   | $\phi_1$ | $\hat{\phi}_1$ | 0.0013 | -0.2284 | 0.0789 | -0.2730 |
|   |   | Bias | 0.7013 | 0.4716 | 0.7789 | 0.4270 |
|   |   | MSE | 0.9785 | 0.2699 | 1.0454 | 0.2211 |
|   |   | SE | 3.8399 | 0.1041 | 0.3321 | 0.1633 |
|   |   | CIL | -0.0391 | -0.2299 | 0.0054 | -0.2899 |
|   |   | CIU | 0.0417 | -0.2269 | 0.1524 | -0.2560 |
|   | $\phi_2$ | $\hat{\phi}_2$ | -0.3809 | 0.2138 | -0.3230 | 0.1971 |
|   |   | Bias | -0.7013 | 0.0938 | -0.4430 | 0.0771 |
|   |   | MSE | 0.9788 | 0.0449 | 1.4744 | 0.0450 |
|   |   | SE | 0.3777 | 0.1455 | 0.9165 | 0.1320 |
|   |   | CIL | -0.5184 | 0.1850 | -0.5298 | 0.1826 |
|   |   | CIU | -0.2434 | 0.2427 | -0.1163 | 0.2116 |
|   | $\sigma$ | $\hat{\sigma}$ | 23.1810 | 1.0546 | 22.1079 | 1.2178 |
|   |   | Bias | 22.1810 | 0.0546 | 21.1079 | 0.2178 |
|   |   | MSE | 557.262 | 0.0189 | 592.552 | 0.0812 |
|   |   | SE | 3.3555 | 0.1666 | 3.3084 | 0.2140 |
|   |   | CIL | 22.2536 | 1.0101 | 21.2234 | 1.1584 |
|   |   | CIU | 24.1085 | 1.0991 | 22.9924 | 1.2771 |
|   |   |   | Normal Error | | $t$ Error | |
|   |   |   | Normal | $t$ | Normal | $t$ |
| 100 | $\beta_1$ | $\hat{\beta}_1$ | 0.3444 | 0.0977 | 0.5861 | 0.1064 |
|   |   | Bias | 0.2444 | -0.0023 | 0.4861 | 0.0064 |
|   |   | MSE | 6.6409 | 0.0164 | 7.8438 | 0.0199 |
|   |   | SE | 5.4338 | 0.3981 | 4.7214 | 0.4410 |
|   |   | CIL | -0.2259 | 0.0287 | 0.3555 | 0.0258 |
|   |   | CIU | 0.9146 | 0.1668 | 0.8168 | 0.1869 |
|   | $\beta_2$ | $\hat{\beta}_2$ | 0.5080 | 0.5049 | 0.9318 | 0.4921 |
|   |   | Bias | 0.0080 | 0.0049 | 0.4318 | -0.0079 |
|   |   | MSE | 8.6819 | 0.0164 | 8.2327 | 0.0310 |
|   |   | SE | 8.2289 | 0.4066 | 4.1940 | 0.3476 |
|   |   | CIL | -0.2487 | 0.4306 | -0.6039 | 0.4320 |
|   |   | CIU | 1.2647 | 0.5793 | 2.4675 | 0.5521 |
|   | $\beta_3$ | $\hat{\beta}_3$ | 0.9028 | 0.9085 | 0.6209 | 0.9054 |
|   |   | Bias | 0.0028 | 0.0085 | -0.2791 | 0.0054 |
|   |   | MSE | 5.9380 | 0.0162 | 11.6861 | 0.0278 |
|   |   | SE | 9.1633 | 0.2197 | 5.5664 | 0.1939 |
|   |   | CIL | -0.0449 | 0.8551 | -1.4063 | 0.8350 |
|   |   | CIU | 1.8505 | 0.9619 | 2.6481 | 0.9759 |

|  |  |  |  |  |  |
|---|---|---|---|---|---|
| $\phi_1$ | $\hat{\phi}_1$ | 0.0245 | -0.1722 | -0.0047 | -0.2039 |
|  | Bias | 0.7245 | 0.5278 | 0.6953 | 0.4961 |
|  | MSE | 0.6194 | 0.3110 | 0.5554 | 0.2652 |
|  | SE | 0.7856 | 0.0835 | 0.0976 | 0.0734 |
|  | CIL | 0.0101 | -0.1792 | -0.0167 | -0.2183 |
|  | CIU | 0.0389 | -0.1653 | 0.0073 | -0.1894 |
| $\phi_2$ | $\hat{\phi}_2$ | -0.1095 | 0.2172 | -0.0639 | 0.2186 |
|  | Bias | -0.2295 | 0.5278 | -0.1839 | 0.0986 |
|  | MSE | 0.1745 | 0.0259 | 0.1300 | 0.0281 |
|  | SE | 0.3773 | 0.0513 | 0.0244 | 0.0555 |
|  | CIL | -0.1226 | 0.2068 | -0.0763 | 0.1989 |
|  | CIU | -0.0965 | 0.2275 | -0.0515 | 0.2383 |
| $\sigma$ | $\hat{\sigma}$ | 27.4424 | 1.1533 | 28.8648 | 1.3182 |
|  | Bias | 26.4424 | 0.1533 | 27.8648 | 0.3182 |
|  | MSE | 770.938 | 0.0422 | 755.250 | 0.1214 |
|  | SE | 2.9298 | 0.1340 | 2.7141 | 0.1875 |
|  | CIL | 26.8991 | 1.1269 | 28.2933 | 1.2838 |
|  | CIU | 27.9858 | 1.1796 | 29.4363 | 1.3526 |

Table 2 shows the simulation results with 10 percent outlier in the data. When outliers are introduced in the data the estimators based on normal distribution are drastically worsen which is reflected to the higher MSE values. However, the estimators based on $t$ distribution still have excellent performance with outliers. The estimators based on $t$ distribution superior to the estimators of the normal distribution in terms of MSE and bias values.

**4.2 Real data examples**

*Example 1*. In this example we will analyze the data set given by Sheather [16]. The data shows that Australian Film Commission's (ACF) yearly gross box office receipts from movies screened in Australia. Table 3 shows the data set. In that book two different scenarios have been applied to this data set. The first one is the ordinary regression model and the second one is the regression model with autoregressive error terms. In this paper we also consider two models and estimate the parameters of interest using the normal and $t$ distributions. Table 4 shows the summary of the estimates, standard errors and the 95% confidence intervals for $\hat{\beta}$ and $\hat{\phi}$. We calculated the standard errors and the 95% confidence intervals using the observed Fisher information.

Table 3. Australian Gross Box Office Results from 1976 to 2007

| Gross box office ($M) | Year | Gross box office ($M) | Year |
|---|---|---|---|
| 95.3 | 1976 | 334.3 | 1992 |
| 86.4 | 1977 | 388.7 | 1993 |
| 119.4 | 1978 | 476.4 | 1994 |
| 124.4 | 1979 | 501.4 | 1995 |
| 154.2 | 1980 | 536.8 | 1996 |
| 174.3 | 1981 | 583.9 | 1997 |
| 210 | 1982 | 629.3 | 1998 |
| 208 | 1983 | 704.1 | 1999 |
| 156 | 1984 | 689.5 | 2000 |
| 160.6 | 1985 | 812.4 | 2001 |
| 188.6 | 1986 | 844.8 | 2002 |
| 182.1 | 1987 | 865.8 | 2003 |
| 223.8 | 1988 | 907.2 | 2004 |
| 257.6 | 1989 | 817.5 | 2005 |
| 284.6 | 1990 | 866.6 | 2006 |
| 325 | 1991 | 895.4 | 2007 |

We consider the following model

$$GrossBoxOffice = \beta_0 + \beta_1 Years + e_t$$

and we assume that the error terms have AR(1) model. The LS estimates show that $\beta_0$ is insignificant so that $\beta_0$ is not included in the model. The LS estimates are used as the initial values for the algorithms.

Table 4. Parameter estimates for Australian Gross Box Office Results

| | | Normal | t |
|---|---|---|---|
| $\beta_1$ | $\hat{\beta}_1$ | 27.1927 | 26.7505 |
| | SE | 0.4263 | 0.7030 |
| | 95% CI | (27.0449) – (27.3404) | (26.5069) – (26.9941) |
| $\phi$ | $\hat{\phi}$ | 0.8816 | 0.2558 |
| | SE | 2.9578 | 0.0066 |
| | 95% CI | (-0.1436) – (1.9067) | (0.2358) – (0.2817) |
| | AIC | 344.2041 | 344.2835 |
| | BIC | 347.1356 | 347.2149 |

Table 4 gives a summary of estimation with both cases. The table shows the estimates standard errors and 95% confidence intervals for two different estimation methods.

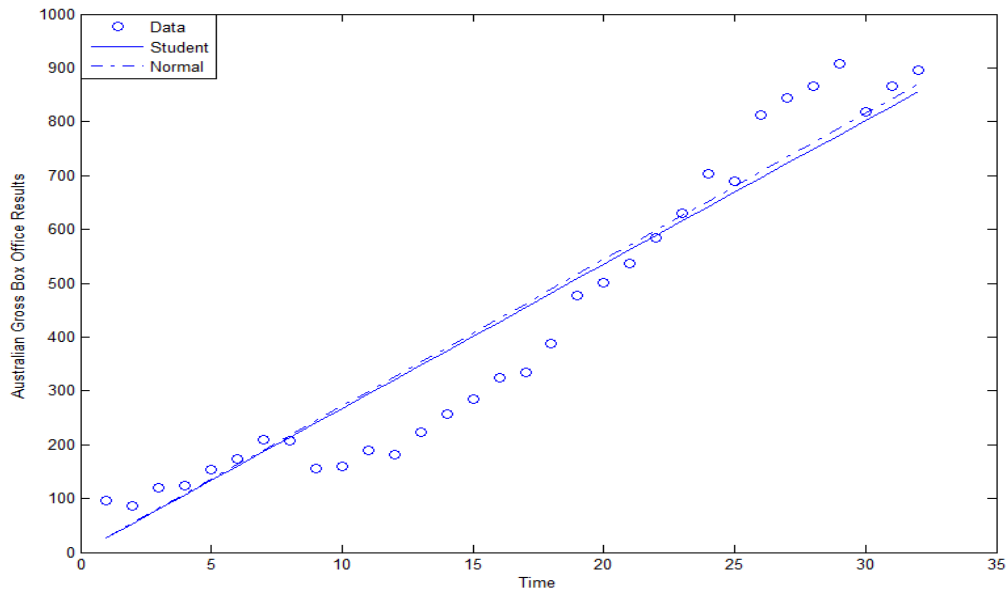

Figure 1. Australian Gross Box Office Results data set estimated by *t* and normal distributions methods

Figure 1 shows the fitted regression lines on the data. The solid line shows the fitted regression obtained from the *t* distribution and the dashed line corresponds to the normal case. When data has no outlier the fitted lines obtained from the normal and the *t* distributions have similar behavior.

*Example 2.* In this example we will analyze the data set given by Rousseeuw and Leroy [15]. The data shows the proportion of the number of ten million international phone calls from Belgium in the years 1950-1973. From Figure 2 we can observe that there are outliers in the data. Rousseeuw and Leroy [15] modeled this data set with a linear regression model to illustrate the performance of the robust regression method the least median of squares (LMS). The following table displays the data.

Table 5. Number of International Calls from Belgium

| Number of Calls[a] ($y_i$) | Year ($x_i$) | Number of Calls[a] ($y_i$) | Year ($x_i$) |
|---|---|---|---|
| 0.44 | 50 | 1.61 | 62 |
| 0.47 | 51 | 2.12 | 63 |
| 0.47 | 52 | 11.90 | 64 |
| 0.59 | 53 | 12.40 | 65 |
| 0.66 | 54 | 14.20 | 66 |
| 0.73 | 55 | 15.90 | 67 |
| 0.81 | 56 | 18.20 | 68 |
| 0.88 | 57 | 21.20 | 69 |
| 1.06 | 58 | 4.30 | 70 |
| 1.20 | 59 | 2.40 | 71 |
| 1.35 | 60 | 2.70 | 72 |
| 1.49 | 61 | 2.90 | 73 |

[a] In tens of millions.

We observe that the OLS residuals show an autocorrelated structure with type AR(1). This observation based on autocorrelation function and the partial autocorrelation function graphs of the OLS residuals. Therefore, we use a regression model with autoregressive error term with AR(1) to model this data set and use normal and the $t$ distribution to obtain estimates for the parameters. The following table gives the summary of the estimates along with the standard errors and the 95% confidence intervals. We also provide the values of the AIC and BIC criteria. The values of AIC and BIC show that the $t$ distribution gives the better fit then the normal distribution. It should be noticed that the estimates obtain from $t$ distribution are much closed to the values obtained from the LMS.

Table 6. Parameter Estimates for International Calls From Belgium

|   |   |   | Normal | $t$ |
|---|---|---|---|---|
| $\beta_0$ |   | $\hat{\beta}_0$ | -13.8142 | -5.3724 |
|   |   | SE | 10.9242 | 11.6686 |
|   |   | 95% CI | (-18.1848) – (-9.4436) | (-10.0408) – (-0.7040) |
| $\beta_1$ |   | $\hat{\beta}_1$ | 0.2980 | 0.1131 |
|   |   | SE | 0.1796 | 0.1918 |
|   |   | 95% CI | (0.2262) – (0.3699) | (0.0363) – (0.1898) |
| $\phi$ |   | $\hat{\phi}$ | 0.7366 | 0.1627 |
|   |   | SE | 50.0647 | 0.2422 |
|   |   | 95% CI | (-19.2934) – (20.7667) | (0.0658) – (0.2596) |
|   | AIC |   | 61.3656 | 37.4305 |
|   | BIC |   | 120.2654 | 72.3951 |

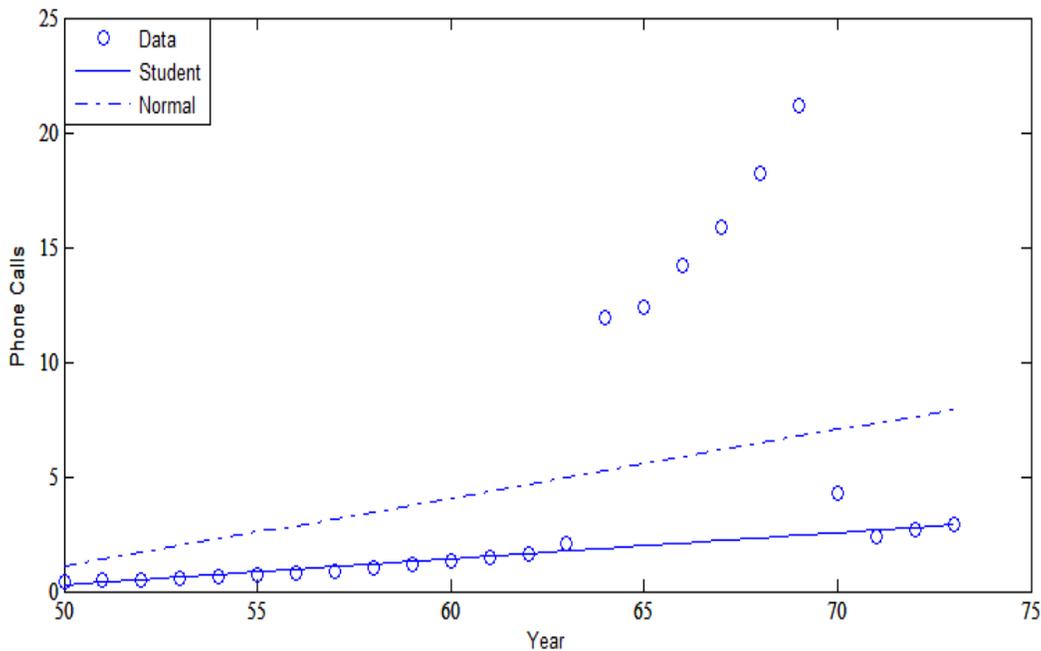

Figure 2. Number of International Phone Calls from Belgium Data Set Estimated by $t$ and Normal Distributions

Figure 2 depicts the scatter plot of the data with the fitted regression lines obtain from normal and the *t* distributions. From this figure we observe that unlike the fitted line obtain from the normal distribution, the fitted line from the *t* distribution is not affected from the outliers.

*Example 3.* We use the data set has been previously analyzed by Ramanathan [14] to show the performance of autoregressive error terms regression model. The data provides consumption of electricity by residential customers served by San Diego Gas and Electric Company. This data set consists of 87 quarterly observations for each 4 covariates from the second quarter of 1972 through fourth quarter of 1993. The response variable is electricity consumption as measured by the logarithm of the kwh (LKWH) sales per residential customer. The explanatory variables are the per-capita income (LY), the price of electricity (LPRICE), cooling degree days (CCD) and heating degree days (HDD). The linear regression model and the expected signs of the $\beta$'s considered in Ramanathan [14] are as follows:

$$LKWH = \beta_0 + \beta_1 LY + \beta_2 LPRICE + \beta_3 CDD + \beta_4 HDD + \varepsilon_t$$

$$\beta_1 > 0, \ \beta_2 < 0, \beta_3 > 0, \beta_4 > 0.$$

It is pointed out by Ramanathan [14] that when the OLS method is used to obtain the estimates the signs of LPRICE, CDD and HDD are consistent with the expected ones, but estimation of LY has the reverse sign. They note that this unexpected result may happen due to ignoring the autocorrelation structure of the error term, hence they suggest using autoregressive error term regression model to model the data. They select the AR order 4 which minimizes the BIC. Then, they used the OLS method to find the estimators for the parameters, and they observed that the sign of the LY is changed towards the expectations.

Here we use the normal and the *t* distributions as the error distribution and obtain the estimators for the parameters of interest. In Table 7 we give the estimates and the AIC values obtained from the normal and the *t* cases. Note that the estimates obtained from the normal distribution are the same with the OLS estimates for the AR(4) model given in Ramanathan [14].

Table 7. The estimated coefficients without outlier

|         |             | OLS      | LS       | LASSO    | Bridge   | MMLASSO  | MMBridge |
|---------|-------------|----------|----------|----------|----------|----------|----------|
| **LY**  | $\hat{\beta}_1$ | **-0.00234** | 0.18625  | 0.05879  | -        | 0.14879  | -        |
| LPRICE  | $\hat{\beta}_2$ | -0.01856 | -0.09354 | -0.08455 | -0.08563 | -0.06455 | -0.08563 |
| CDD     | $\hat{\beta}_3$ | 0.06365  | 0.00029  | 0.00028  | 0.00028  | 0.00028  | 0.00028  |
| HDD     | $\hat{\beta}_4$ | 0.08564  | 0.00022  | 0.00022  | 0.00023  | 0.00022  | 0.00023  |
| AR order |            | -        | 4        | 4        | 4        | 4        | 4        |

Table 7. The estimated coefficients with outlier

|         |             | OLS      | LS       | LASSO    | Bridge   | MMLASSO  | MMBridge |
|---------|-------------|----------|----------|----------|----------|----------|----------|
| **LY**  | $\hat{\beta}_1$ | **-2.69756** | 1.69845  | 3.65769  | -        | 0.25654  | -        |
| LPRICE  | $\hat{\beta}_2$ | -0.96123 | -0.00154 | -0.98555 | -0.64786 | -0.00547 | -0.02645 |
| CDD     | $\hat{\beta}_3$ | 0.57743  | 0.07264  | 0.64135  | 0.91231  | 0.00072  | 0.00036  |
| HDD     | $\hat{\beta}_4$ | 0.96874  | 0.04622  | 0.95344  | 0.84521  | 0.00095  | 0.00041  |
| AR order |            | -        | 4        | 4        | 4        | 4        | 4        |

We notice from this table that unlike the OLS estimate without AR(4) structure the sign of $\beta_1$ is positive when the autoregressive errors are assumed as Ramanathan [14] reported. Further, when the estimation is carried out using the *t* distribution, the AIC value is smaller than the AIC obtained from the normal distribution. Thus, it can be concluded that the *t* distribution may provide better fit than the normal distribution for this data.

**5. Discussion**

In this paper, we have proposed to use the *t* distribution as an alternative to the normal distribution as the error distribution in linear regression model with autoregressive error terms. The simulation results and the real data examples have shown that the *t* and the normal distributions give similar results when there is no outlier in the data. On the other hand, when the data have some outliers the *t* distribution has better performance than the normal distribution for all the settings. Further, Example 3 has shown that the OLS method may fail to accurately estimate the unknown parameters when the error terms have autocorrelation structure. However, when the autoregressive error form is introduced into the model the estimates are correctly obtained in terms of sign from the OLS method, the normal and the *t* distributions. For the same example we have also noticed that the result obtained from the *t* distribution is better than the result obtained from the normal according to AIC values, which may show that the *t* distribution can be better model than the normal distribution. To sum up, all of these results show that the *t* distribution can be used as an alternative to the normal distribution for parameter estimation in a linear regression model with autoregressive error terms when the data sets have outliers and/or heavy tailed error distributions.

**Acknowledgements**

The authors thank the anonymous referee, the editor and the associate editor whose comments, suggestions, and corrections have led to a considerably improvement of this paper.

**References**

[1]  Ansley, C. F. (1979). An algorithm for the exact likelihood of a mixed autoregressive-moving average process. Biometrika, 66 (1): 59-65.

[2]  Alpuim, T. and El-Shaarawi, A. (2008). On the efficiency of regression analysis with AR(p) errors. Journal of Applied Statistics, 35:7, 717-737.

[3]  Arslan, O. and Genç, A. İ. (2003). Robust location and scale estimation based on the univariate generalized *t* (GT) distribution. Communications in Statistics – Theory and Methods, 32:8, 1505-1525.

[4] Arslan, O. and Genç, A. İ. (2009). The skew generalized *t* distribution as the scale mixture of a skew exponential power distribution and its applications in robust estimation. Statistics: A Journal of Theoretical and Applied Statistics, 43:5, 481-498.

[5] Beach, C.M. and McKinnon, J. G. (1978). A maximum likelihood procedure for regression with autocorrelated errors. Econometrica, 46:1, 51- 58.

[6] Huang, X., Zhou, Y. and Zhang R. (2004). A multiscale model for MPEG-4 varied bit rate video traffic. IEEE Transactions on Broadcasting, 50:3, 323-334.

[7] Hill, J. B. (2013). Least tail-trimmed squares for infinite variance autoregressions. Journal of Time Series Analysis, 34, 168-186.

[8] Hill, J. B. (2015a). Robust estimation and inference for heavy tailed GARCH. Bernoulli, 21, 1629-1669.

[9] Hill, J. B. (2015b). Robust generalized empirical likelihood for heavy tailed autoregressions with conditionally heteroscedastic errors. Journal of Multivariate Analysis, 135, 131-152.

[10] Lucas, A. (1997). Robustness of the student t based M-estimator. Communications in Statistics – Theory and Methods, 26(5), 1165-1182.

[11] Lange, K.L., Little, R.J.A. and Taylor, J.M.G. (1989). Robust statistical modeling using the *t*-distribution. J. Am. Stat. Assoc., 84, 881–896.

[12] McLachlan, G., Krishnan, T. (1997). The EM Algorithm and Extensions. Wiley Series in Probability and Statistics, USA.

[13] Olaomi, J.O. and Ifederu, A. (2008). Understanding estimators of linear regression model with AR(1) error which are correlated with exponential regressor. Asian Journal of Mathematics and Statistics, 1(1), 14-23.

[14] Ramanathan, R. (1998). Introductory Econometrics with Applications. Fort Worth: Dryden: Harcourt Brace College Publishers.

[15] Rousseeuw, P.J. and Leroy, A.M. (1987). Robust Regression and Outlier Detection. Wiley Series, USA.

[16] Sheather S.J. (2009). A Modern Approach to Regression with R. Springer-Verlag.

[17] Tiku, M.L., Wong, W., Bian, G. (1999). Estimating parameters in autoregressive models in non-normal situations: symmetric innovations. Communications in Statistics Theory and Methods, 28(2), 315-341.